\documentclass[journal,12pt,onecolumn, draftclsnofoot]{IEEEtran}

\fontsize{12}{10}

\usepackage[utf8]{inputenc}
\usepackage{times}

\normalsize
\DeclareMathAlphabet\mathbfcal{OMS}{cmsy}{b}{n}

\usepackage[english]{babel}
\usepackage{amsmath,amssymb}
\usepackage{graphicx}
\usepackage{dsfont}
\usepackage{amsthm}
\usepackage{amssymb}
\usepackage{mdframed}
\usepackage{caption}
\usepackage{subcaption}

\usepackage{algorithm}
\usepackage{algorithmicx}
\usepackage{algcompatible}
\usepackage{eqparbox}
\usepackage{enumerate}

\makeatletter
\def\blfootnote{\xdef\@thefnmark{}\@footnotetext}
\makeatother

\usepackage[x11names]{xcolor}
\newif\ifrevision
\newcommand{\highlight}[1]{%
\ifrevision
\textcolor{red}{#1}%
\else
#1%
\fi
}

\usepackage{lscape}
\usepackage{bbm}

\newtheorem*{previous theorem}{Theorem~1 of \cite{journal_paper}}

\title{Communicate to Learn at the Edge}

\author{
\IEEEauthorblockN{\normalsize D. G{\"u}nd{\"u}z$^1$, D. Burth Kurka$^1$, M. Jankowski$^1$, \\ M. Mohammadi Amiri$^2$, E. Ozfatura$^1$, and S. Sreekumar$^3$}\\
\IEEEauthorblockA{\small $^1$Department of Electrical and Electronic Engineering, Imperial College London}\\
\IEEEauthorblockA{\small $^2$Department of Electrical Engineering, Princeton University}\\
\IEEEauthorblockA{\small $^3$Department of Electrical and Computer Engineering, Cornell University}
}

\begin{document}

\revisionfalse 

\bstctlcite{IEEEexample:BSTcontrol}

\maketitle

\begin{abstract}
Bringing the success of modern machine learning (ML) techniques to mobile devices can enable many new services and businesses, but also poses significant technical and research challenges.  Two factors that are critical for the success of ML algorithms are massive amounts of data and processing power, both of which are plentiful, yet highly distributed at the network edge. Moreover, edge devices are connected through bandwidth- and power-limited wireless links that suffer from noise, time-variations, and interference. Information and coding theory have laid the foundations of reliable and efficient communications in the presence of channel imperfections, whose application in modern wireless networks have been a tremendous success. However, there is a clear disconnect between the current coding and communication schemes, and the ML algorithms deployed at the network edge. In this paper, we challenge the current approach that treats these problems separately, and argue for a joint communication and learning paradigm for both the training and inference stages of edge learning.
\end{abstract}

\blfootnote{This work was supported by the European Research Council (ERC) Starting Grant BEACON (grant agreement no. 677854).}


\section{Motivation}

Modern machine learning (ML) techniques have made tremendous advances in areas such as machine vision, robotics, and natural language processing. Novel ML applications emerge every day, ranging from autonomous driving and finance to marketing and healthcare – potential applications are limitless. In parallel, the fifth generation (5G) of mobile technology promises to connect billions of heterogeneous devices to the network edge, supporting new applications and verticals under the banner of Internet of things (IoT). Edge devices will collect massive amounts of data, opening up new avenues for ML applications. The prevalent approach for the implementation of ML solutions on edge devices is to amass all the relevant data at a cloud server, and train a powerful ML model using all the available data and processing power. However, such a `centralized' solution is not applicable in many cases. This might violate the latency requirements of the underlying application, particularly in the inference stage; or, result in the infringement of user privacy.
Moreover, as the data volumes increase, limited bandwidth and energy resources of IoT devices will become a bottleneck. For example, an autonomous car generates 5 to 20 terabytes of data per day. This is a particular challenge when the `information density' of the collected data is low, i.e., large volumes of data with only limited relevant information for the underlying learning task.

To meet the requirements of most IoT applications, the `intelligence’ should move from the centralized cloud to the network edge. However, both data and processing power, the essential constituents of machine intelligence, are highly distributed at the edge. As a result, \textbf{communication becomes key to an intelligent network edge}, and potential solutions must allow edge devices not only to share their data but also computational resources in a seamless and efficient manner. We can argue that the current success of ML, thanks to the tremendous increase in computational power, is similar to the `great leap forward' in human evolution, which led to the development of human brain thanks to a favorable mutation. Continuing with this analogy, next big revolution in ML is likely to arrive thanks to the efficient orchestration and collaboration among intelligent devices, similarly to the impact of language in human history, which tremendously accelerated the advancement of our civilization by allowing humans to share information, experience, and intelligence.

\begin{figure}[t]
  \begin{center}
    \includegraphics[width=0.5\textwidth]{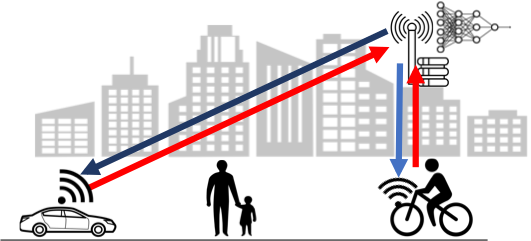}
  \end{center}
  \caption{Distributed learning and inference at the wireless network edge.}
  \label{f:distr_inference}
\end{figure}

\subsection{The Communication Challenge}

Communication bottleneck in ML has been acknowledged in the literature; yet, most current approaches treat communication links as rate-limited ideal bit pipes. However, wireless links introduce errors due to noise and channel fading, and error-free operation is either impossible, or would result in significant delays. This is particularly prominent at the network edge, where bandwidth- and power-limited IoT devices share the same wireless medium, also creating interference to each other. Moreover, when information moves across a network, privacy and security concerns arise, exacerbated at the edge due to the vulnerability of individual devices and the broadcast nature of wireless transmissions. %

After decades of research, communication engineers have designed highly advanced coding and communication techniques that can mitigate channel imperfections and create reliable links among wireless devices; however, reducing the communication among edge devices to a network of ideal bit pipes has the following limitations: 1) communication protocols that enable such reliable links introduce significant overheads and delays, which are not acceptable for many ML applications; 2) such levels of reliability at the link level may not be required for some ML applications, resulting in inefficient resource management; 3) most communication protocols are designed to reduce or remove interference, which may not be desired in some distributed ML applications. \textbf{To overcome these limitations, we need to reconsider physical layer and networking solutions taking into account the limitations and requirements of the underlying ML applications.} 

Information and coding theory have laid the foundations of reliable, efficient and secure communication in the presence of channel imperfections and interference, whose application in modern wireless networks have been a tremendous success. While the fundamental information theoretic ideas and coding theoretic tools can play an important role in enabling fully distributed learning across distributed heterogeneous edge devices, many of the existing concepts and techniques are not relevant for ML applications, whose communication requirements and constraints (latency, reliability, security, privacy, etc.) are fundamentally different from the type of traffic current networks are designed for. Moreover, as we will try to show in this paper, we cannot overcome these limitations by a simple `cross-layer' approach, i.e., by tuning the parameters of existing communication protocols. There is a clear disconnect between the current coding and communication techniques, and the ML algorithms and architectures that must be deployed at the network edge, and \textbf{we need a fundamentally new paradigm of coding, communication and networking with ML applications in mind.}

Next, we present the challenges in achieving a fully distributed edge intelligence across heterogeneous agents communicating over imperfect wireless channels. We will treat the \textit{inference} and \textit{training} phases of ML algorithms separately as they have distinct reliability and latency requirements. 

\noindent
\section{Distributed Inference}\label{s:inference} 

\textit{Inference} refers to applying a trained model on a new data sample to make a prediction. Although inference tasks require much less computational resources compared to training, they typically impose more strict latency constraints. For example, in self-driving cars (see Fig.~\ref{f:distr_inference}), immediate detection of obstacles is critical to avoid accidents. A powerful deep neural network (DNN) model can be pre-trained and deployed for this task. However, it is often not possible to carry out inference locally at a single device, as decisions may rely on data (e.g., background and terrain information) available at an edge server, or on signals from other cars; or the device gathering the data (e.g., a bike) may not have the necessary processing capability. Communication becomes indispensable in such scenarios, and we need to guarantee that 
inference can still be accomplished within the accuracy and latency constraints of the underlying application. 

\begin{figure}[t]
  \begin{center}
    \includegraphics[width=0.6\textwidth]{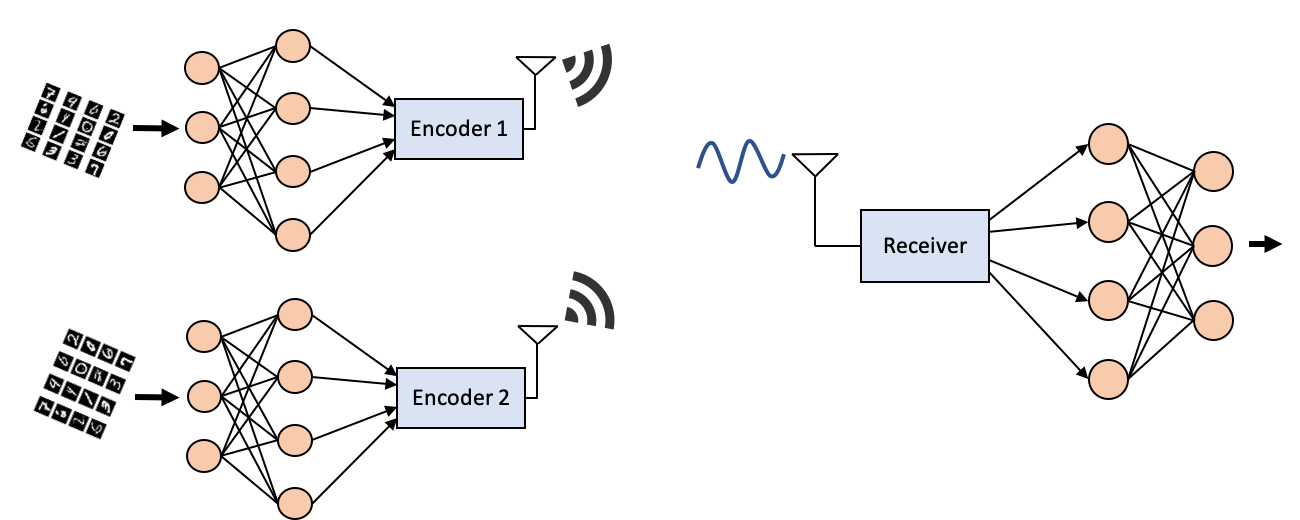}
  \end{center}
  \caption{DNNs at the network edge.}
  \label{f:distr_DNNs}
\end{figure}

\textbf{Fundamental limits.} As a first step towards understanding the fundamental limits of statistical inference over noisy channels, a distributed binary hypothesis testing (HT) problem is studied in \cite{Sreekumar:IT:20}. Consider two devices with their local observations. 
One of the devices (e.g., the car in Fig. \ref{f:distr_inference}), called the \textit{observer}, conveys some information about its observations to the other one, called the \textit{decision maker} (e.g., the edge server in Fig. \ref{f:distr_inference}), over a noisy channel. The decision maker has to make a decision on the joint distribution of the observations of the two devices. Since the observer has access only to its own observations, it cannot make a local decision no matter how much processing power it has; instead it must convey some features of its observations to help the decision maker to make the correct decision. The question here is whether the features and the channel code to transmit them can be designed separately. If the goal were to transmit the samples at the observer with the minimal average distortion (under any additive finite distortion measure), according to Shannon's separation theorem the compression and channel coding tasks can be carried out separately and without loss of optimality, in the limit of infinite blocklength. However, it is shown in \cite{Sreekumar:IT:20} that the optimality of separation breaks down in the remote HT problem, as the goal here is to decide on the joint distribution with minimal error probability. 
While this result shows that communication and inference cannot be separated even in the asymptotic limit (without loss of optimality), how a joint scheme should be designed in practice is a vastly unexplored research direction with great potential in future edge inference applications. \highlight{Next, we provide several practical examples of edge inference problems, and illustrate how jointly treating communication and inference can help improve both the speed and the accuracy of the inference task.}

\textbf{Edge Inference with DNNs.} DNNs achieve the state-of-the-art performance in most ML tasks.
In distributed inference across mobile devices and edge servers, a common approach is to partition a pre-trained DNN baseline between the devices and the edge server depending on the former's computational capabilities (see Fig.~\ref{f:distr_DNNs}) \cite{JointDNN}.  Conventional approaches abstract out the wireless channel as an error-free ideal bit-pipe, and focus only on the feature compression problem, ignoring the potential impacts of communication in terms of delay, complexity, and reliability. However, lossy transmission of feature vectors over a wireless channel is a \textit{joint source-channel coding (JSCC) problem}, and
separation is known to be suboptimal under strict latency constraints imposed by inference problems.

While JSCC has long been studied, mainly for image and video transmission, these works mostly took a model-driven approach exploiting particular properties of the underlying source and channel statistics. Recently, an alternative fully data-driven DNN-based scheme, called DeepJSCC, has been introduced in \cite{deepJSCC}. DeepJSCC not only beats digital alternatives for image transmission (e.g., BPG image compression + LDPC channel coding), but also provides `graceful degradation' with channel quality, making it ideal for IoT applications, where accurate channel estimation is often not possible. 
DeepJSCC also reduces the coding/decoding delay compared to conventional digital schemes more than 5 times on a CPU, and more than 10 times on a GPU. 
As opposed to conventional digital schemes, DeepJSCC can easily adapt to specific information source or channel statistics through training, e.g., landscape images transmitted from a drone or a satellite. This makes DeepJSCC especially attractive for edge inference as we do not have compression codes designed for feature vectors, whose statistics would change from application to application.

\begin{figure}[t]
  \begin{center}
\includegraphics[width=3.3in]{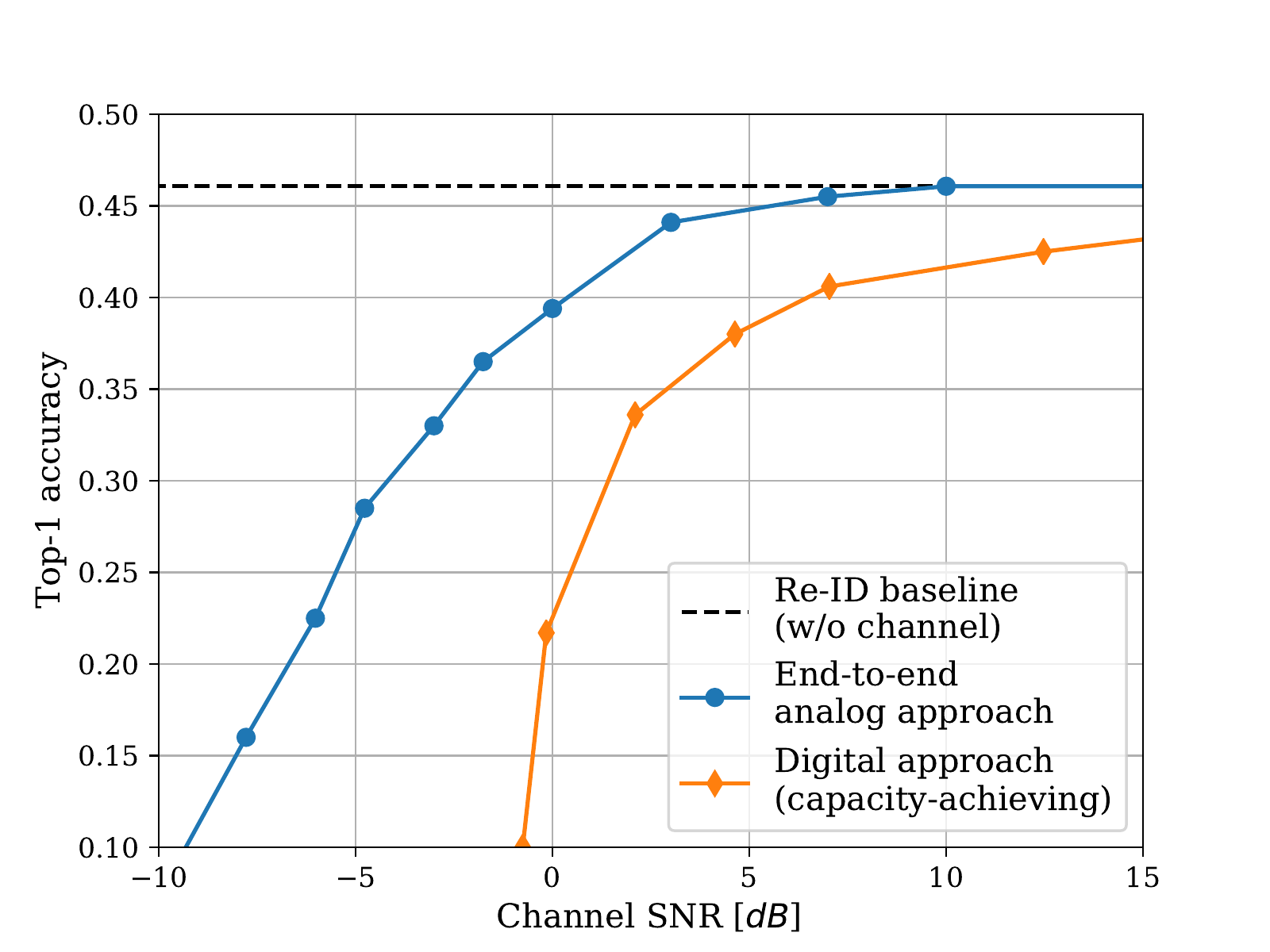}
  \end{center}
\caption{Accuracy vs. channel SNR for remote person re-ID over an AWGN channel.}
\label{f:remote_reID}   
\end{figure}

A practical edge inference problem is studied in \cite{jankowski2019deep}, where the image of a person captured by a remote camera is to be identified within a database available at an edge server, called the \textit{re-identification (re-ID) problem}. Here, the camera cannot make a local decision as it does not have access to the database. In \cite{jankowski2019deep}, two approaches are proposed, both employing DNNs for remote inference: a task-oriented DNN-based compression scheme for digital transmission and a DNN-based analog JSCC approach, {\`a} la DeepJSCC. These schemes are compared in Fig.~\ref{f:remote_reID} in terms of top-1 identification accuracy when only 128 real symbols are transmitted over an additive white Gaussian noise (AWGN) channel. We observe that the analog approach, which maps the feature vectors directly to channel inputs (no explicit compression or channel coding), performs significantly better, achieving the baseline performance around a channel signal-to-noise-ratio (SNR) of approximately 8~dB. We highlight that the conventional scheme of transmitting the query images with the best possible quality (ignoring the learning task), and then applying the re-ID baseline on the reconstructed image is not included as it would require much higher SNR values to achieve a comparable performance. This result shows that \textbf{separating communication from inference at the edge can be highly suboptimal}. While joint design can offer significant performance gains, it brings about new challenges and requires novel coding and communication paradigms, \highlight{including the extension of the proposed edge inference approach to time-varying and/or non-Gaussian channels, and to multi-antenna and multi-user networks}.

\highlight{In the inference stage, the challenge is to convey the most relevant information about the data samples to the decision maker to achieve the desired level of accuracy within the constraints of the edge network. The results above show that the channel characteristics must be taken into account during the training stage, rather than being abstracted out, and effectively, we learn how to communicate and infer jointly. In this section, we have assumed that the DNNs are trained centrally, and then deployed at the edge devices, assuming the availability of sufficient training data and an accurate model of the wireless communication channels. We focus on the training stage in the next section.}

\begin{figure}
     \centering
     \begin{subfigure}{0.7\textwidth}
         \centering
         \includegraphics[width=.7\textwidth]{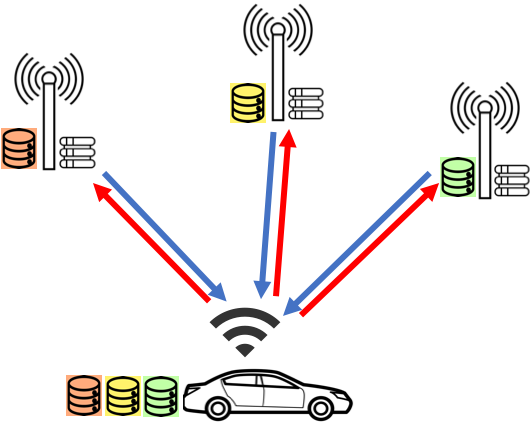}
         \caption{Distributed training with centralized data.}
         \label{f:distr_training}
     \end{subfigure}
     \hfill
     \begin{subfigure}[b]{0.5\textwidth}
         \centering
         \vspace{.4in}
         \includegraphics[width=\textwidth]{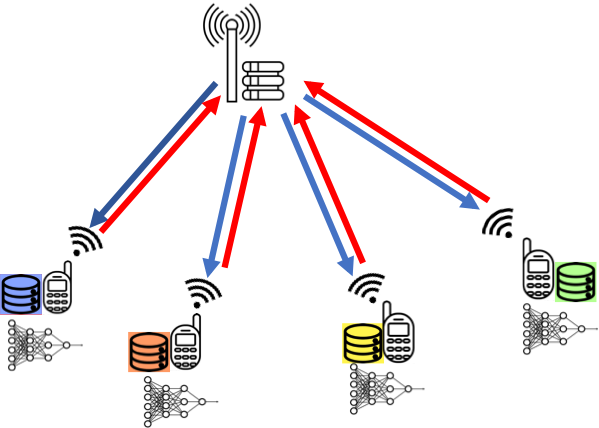}
         \caption{FEEL with distributed data.}
         \label{f:federated_learning}
     \end{subfigure}
        \caption{Distributed training at the edge.}
        \label{fig:three graphs}
\end{figure}

\section{Distributed Training}\label{s:training}

Training is particularly challenging at the network edge due to the distributed nature of both the data and the processing power. Below, we will first address the scenario in which an edge device with its own dataset employs the computational resources of multiple edge servers to speed up training (see Fig.~\ref{f:distr_training}). Later, we will consider the scenario when data is also distributed (see Fig.~\ref{f:federated_learning}).

In the training stage of a standard ML problem, the goal is to optimize the model parameters over a training dataset with respect to an application specific {\em empirical loss function}. This optimization problem is typically solved by stochastic gradient descent (SGD), iteratively updating the parameter vector along the estimated gradient descent direction. This algorithm is highly parallelizable, allowing distributed and parallel implementation. When the dataset is large, distributed SGD  across multiple edge servers can be utilized to reduce the training time. The dataset can be divided into non-overlapping subsets, each given to a different server. At each iteration of the gradient descent algorithm, the user broadcasts the current model parameters to all the servers. Each server computes a partial gradient based only on its local dataset, and returns the result to the master. The master waits to receive partial gradients from all the servers in order to aggregate them and obtain the full gradient. In this implementation, however, due to {\em synchronised updates} the completion time of each iteration is constrained by the \textit{straggling server(s)}, where the straggling may be due to failing hardware, contention in the network, or even channel outages if the training is carried out at the wireless edge.

Straggling servers can be treated as `erasures', and using ideas from coding theory, redundant computations can be introduced to efficiently compensate for erasures \cite{Lee:Code2Learn:18, GradientCoding}. This can help reduce the \textit{recovery threshold}, the minimum number of responsive servers required to complete the computation task, e.g., computing a sufficiently accurate gradient estimate. However, this may require coding the data before offloading to the servers \cite{Lee:Code2Learn:18}, or coding the results of computations at each server \cite{GradientCoding}, and eventually decoding these responses by the user, which introduce additional complexity and delays. Despite the significant research efforts in recent years, optimal coding schemes remain elusive, and there is no comprehensive analysis of end-to-end latency that take into account the communication, coding, and computing delays. 

Moreover, most of the existing techniques suffer from two main drawbacks: the recovery threshold can be reduced by increasing the redundancy; yet, the servers may end up executing more computations than required due to an inaccurate prediction of the straggling behaviour, resulting in {\em over-computation}. Also, most of the existing solutions are designed for persistent stragglers, and partial computations carried out by stragglers are discarded, resulting in {\em under-utilization} of the computational resources. To overcome these limitations, each server can be allowed to send multiple messages during each training iteration \cite{Ozfatura:Entropy:20}, each corresponding to partial computations. This approach will provide additional flexibility for straggler mitigation, resulting in a trade-off between the amount of communication and computation. We highlight that the real performance indicator for these schemes is the average completion time of training, which requires the joint design of the underlying communication protocol and the coded computing scheme employed.

\textbf{Private and secure distributed computation.} Distributed training also introduces privacy and security challenges. Malicious servers can inject false data, while honest but curious servers can exploit user data for purposes beyond computation. 
\highlight{Coded computing, in particular polynomial codes, can provide security and privacy guarantees in addition to straggler mitigation by delivering coded data samples to the computing servers \cite{Lagrange_secure_private}, but the optimal trade-off between the required communication bandwidth between the user and the servers, and the privacy/ security guarantees (in terms of the number of colliding servers) remains an open challenge.}

\section{Federated edge learning (FEEL)}\label{s:FEEL}

\highlight{When multiple edge devices with their own local datasets collaborate to train a joint model, devices may not want to offload their data due to privacy concerns. Yet, unlike in distributed training, data samples at different devices cannot be coded to provide privacy}. Federated learning (FL) has been introduced by Google to enable collaborative training without sharing local datasets \cite{McMahan_ICAIS17}, typically orchestrated by a parameter server (PS) (see Fig.~\ref{f:federated_learning}).
In FL, the PS broadcasts a global model to the devices. Each device runs SGD locally using the current global model. Device updates are aggregated at the PS, and used to update the global model. Communication, again, is a major challenge due to the bandwidth and power limitations of devices. To reduce the communication load, random subsets of devices are selected at each round, and local models are communicated after several local SGD updates. \highlight{Another approach is to reduce the size of the messages communicated between the devices and the PS through compression. This is yet another research challenge where the extensive knowledge in information and coding theory for data compression can make an impact. While initial works have focused on rather simple scalar quantization and sparsification techniques \cite{NIPS2017_6768}, more advanced vector quantization and temporal coding tools exploiting correlations across gradient dimensions or multiple iterations can further reduce the communication load. But, the complexity of such tools must be carefully balanced with the potential gains. }

\highlight{In FEEL, we assume that the training takes place at the network edge across wireless devices within physical proximity; therefore, communication from edge devices to the PS will be limited by the power and bandwidth constraints, interference among devices, and time-varying channel fading.} When the model size is relatively small compared to the size of the dataset, exchanging model parameters rather than data provides another advantage of FEEL. Still, allocation and optimization of channel resources among devices will be essential to improve the learning performance. On the other hand, conventional solutions that maximize throughput do not necessarily translate into better accuracy or faster convergence in FEEL \cite{Quek:FL_schedule, Tran:INFOCOM19}. Moreover, conventional measures based on number of iterations may not be relevant in FEEL, as the wall clock time depend hugely on the communication protocol \cite{Tran:INFOCOM19}. Optimizing the communication protocols for FEEL poses many interesting research challenges; however, most current approaches, motivated by conventional communication systems, consider orthogonal resource allocation with the aim of minimizing interference. 

\textbf{Interference can be a bliss.} In the uplink transmission from the devices, the PS is interested only in the average of the local models. Hence, rather than transmitting individual updates in an orthogonal fashion, signal superposition property of the wireless medium can be exploited to directly convey the sum of the local parameters through \textit{over-the-air computation} \cite{amiri2019federated, amiri2019machine}. This is achieved by all the devices synchronously transmitting their model updates in an uncoded `analog' fashion, which are superposed by the channel.

\begin{figure}[t]
  \begin{center}
    \includegraphics[width=0.65\textwidth]{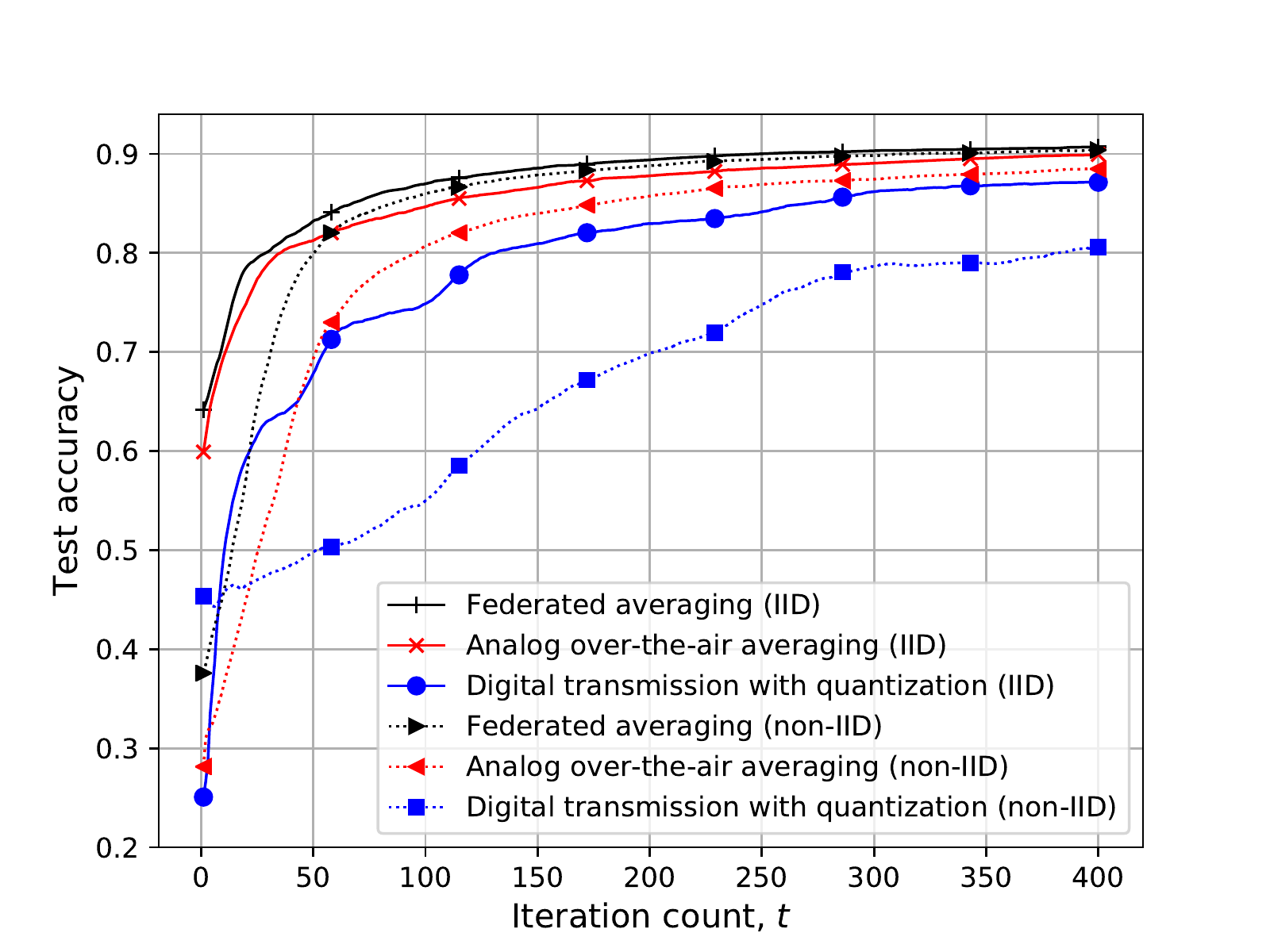}
  \end{center}
  \caption{Test accuracy of FEEL for MNIST classification with IID and non-IID data distributions.}
  \label{f:MNIST_FEEL}
\end{figure}

Uplink transmission of local model updates in FEEL is a distributed computation problem, for which there is no separation theorem even when the sources are independent. Model updates at different devices are highly compressible, and are often correlated. Hence, when model updates are conveyed through digital communication, model compression can be used to adapt to the limited channel resources available to each device \cite{NIPS2017_6768}. In analog transmission, however, even though all the devices transmit over the same channel resources, the required bandwidth can be fairly large. Some state-of-the-art models include tens of millions of parameters, whereas 1 LTE frame of 5MHz bandwidth and 10ms duration can carry only 6K complex symbols. In \cite{amiri2019federated}, sparsification of model updates is proposed followed by linear projection with a pseudo-random Gaussian matrix. This novel approach serves as an analog compression technique, and reliable reconstruction can be achieved by approximate message passing at the PS. In Fig.~\ref{f:MNIST_FEEL}, we compare digital and analog schemes for the MNIST classification task over a Gaussian multiple access channel. In the IID case, local datasets are chosen randomly from the whole training dataset; whereas in the non-IID case each device has samples from only two classes. We see that over-the-air computation provides significant gains in both the final accuracy and the convergence speed. Over-the-air computation allows scheduling more devices within the same time constraint, which provides variance reduction in updates, and better robustness against the channel noise \cite{amiri2019federated}. This is yet another example, where a joint design of the communication and learning algorithms is essential.

\highlight{We remark that over-the-air computation assumes symbol-level synchronization among the participating devices. In practice, this can be achieved through a synchronization channel, e.g., timing advance in LTE systems, resulting in a trade-off between the overall performance and the resources dedicated to synchronization, which is an interesting research direction to fully evaluate the potential benefits of over-the-air computation for FEEL.}

\textbf{Privacy in FEEL.} Although FL has been introduced as a privacy-aware solution for collaborative learning, it is known to be vulnerable to membership as well as reconstruction attacks solely using the gradient information \cite{deep_leakage}. 
Although differential privacy can be achieved by introducing noise into the gradients transmitted by the devices, this typically requires adding significant amount of noise, making the model hard to converge. On the other hand, in FEEL, there is inherent noise and interference in the channel, which can be exploited to increase the security and privacy of the system through purely physical layer techniques. This opens up a new type of physical layer security/ privacy framework for FEEL applications.

\section{Discussion and Conclusions}

Communication will play an essential role in employing ML tools at the network edge. 
Current approaches to communication-efficient distributed ML ignore the physical layer, and assume error and delay-free ideal links. This approach presumes a communication protocol, designed independently of the learning task, taking care of channel imperfections. In this paper, we have argued through references to recent theoretical results and practical implementations that such a separate architecture can be highly suboptimal, and a novel joint communication and learning framework is essential in approaching the fundamental limits of distributed learning. This calls for a new research paradigm integrating coding and communication theoretic ideas within the design of ML algorithms at the network edge. \highlight{We have shown that the benefits of such a joint design paradigm can be significant for edge inference, both to boost the final performance and to meet the stringent delay constraints. Training is more computation intensive compared to inference; hence, computation and communication delays in training must be optimized jointly. Furthermore, heterogeneity of edge servers may result in additional bottlenecks due to stragglers. Coding can be used both to reduce the computation delays and to mitigate stragglers. Moreover, each iteration of the training process can be considered as a distributed  computation  problem, which  renders throughput-maximizing conventional communication protocols obsolete, and requires the design of novel  communication  protocols  and coding schemes. Since training is carried out in many (imperfect) iterations, we can relax some of the constraints  of traditional coding and communication schemes (reliability, synchronization, power control, etc), resulting in novel communication problems. Finally, taking into account the physical layer channel characteristics can allow exploiting coding and communication theoretic tools to provide fundamental information theoretic privacy and security guarantees for both inference and training at the edge.}
Each of these perspectives and challenges open up new research problems in this exciting new research area exploring the connections between communication and learning.

\bibliography{IEEEabrv, test.bib}

\end{document}